\documentclass[runningheads,a4paper]{llncs}

\usepackage{amssymb}
\usepackage{graphicx}
\usepackage{listings}
\usepackage{color}
\usepackage{balance}
\usepackage[utf8]{inputenc}
\usepackage[T1]{fontenc}
\usepackage{microtype}
\usepackage{cite}
\usepackage{url}
\usepackage{tabularx}
\usepackage{booktabs}
\usepackage{multirow}
\usepackage{url}
\usepackage[colorinlistoftodos]{todonotes}
\usepackage{lmodern}
\usepackage{tabularx}
\usepackage{listings}
\usepackage{longtable}

\definecolor{name}{rgb}{0.5,0.5,0.5}
\definecolor{javared}{rgb}{0.6,0,0} 
\definecolor{javagreen}{rgb}{0.25,0.5,0.35} 
\definecolor{javapurple}{rgb}{0.5,0,0.35} 
\definecolor{javadocblue}{rgb}{0.25,0.35,0.75} 

\usepackage{url}
\urldef{\mailsa}\path|{amelie.gyrard, martin.serrano}@insight-centre.org|
\urldef{\mailsb}\path|pankesh.patel@in.abb.com|
\urldef{\mailsc}\path|d.thakker@bradford.ac.uk|
\urldef{\mailsd}\path|amit@knoesis.org|

\begin{document} 

\newcommand{\fakeparagraph}[1]{\smallskip\noindent\textbf{#1.}}

\mainmatter 

\title{SWoTSuite: A Development Framework for Prototyping Cross-domain Semantic Web of Things Applications}

\titlerunning{SWoTSuite: A Toolkit for Prototyping Cross-domain SWoT Applications}

\author{Pankesh Patel\inst{1}
\and Amelie Gyrard\inst{2} \and Dhavalkumar Thakker\inst{3} \and Amit Sheth\inst{4} \and Martin Serrano\inst{2} }
\authorrunning{Demonstration at ISWC 2016}
\institute{ABB Corporate Research, India\\ \mailsb
\and
Insight Center for Data Analytics, National University of Galway, Ireland\\
\mailsa
\and
University of Bradford, Bradford, United kingdom\\ \mailsc
\and
Kno.e.sis and Wright State University, USA\\ \mailsd
}

\toctitle{Demonstration abstract ISWC 2016}
\tocauthor{Authors' Instructions}
\maketitle

\begin{abstract}
Semantic Web of Things~(SWoT) applications focus on providing a wide-scale interoperability that allows the sharing of IoT devices across domains and the reusing of available knowledge on the web. However, the application development is difficult because developers have to do various tasks such as designing an application, annotating IoT data, interpreting data, and combining application domains. 

To address the above challenges, this paper demonstrates SWoTSuite, a toolkit for prototyping SWoT applications. It hides the use of semantic web technologies as much as possible to avoid the burden of designing SWoT applications that involves designing ontologies, annotating sensor data, and using reasoning mechanisms to enrich data. Taking inspiration from  sharing and reuse approaches, SWoTSuite reuses data and vocabularies. It leverages existing technologies to build applications. We take a hello world naturopathy application as an example and demonstrate an application development process using SWoTSuite. The demo video is available at URL \url{http://tinyurl.com/zs9flrt}.
\end{abstract}

\section{Introduction}\label{sec:intro}

In recent years, we have been witnessing a growing number of applications exploiting sensors are becoming increasingly popular. However, existing applications are largely specific to a domain and they lack methods to create innovative cross-domain applications. Some examples of cross-domain SWoT applications~\cite{gyrard:tel-01217561} include recommending food according to  weather forecasts, home remedies according to health measurements, and  safety equipment in a smart car according to the weather. To build these applications, developers have to perform various tasks such as designing an application, semantically annotating IoT data, and interpreting IoT data. This requires the learning of  semantic web technologies and tools, which is a time consuming process. 

The aim of our work is to reduce the development effort for prototyping SWoT 
applications~\cite{gyrard:tel-01217561, Chauhan:2016:DFP:2897035.2897039, 7460669}. Taking inspiration from Content Management Systems~(e.g., Drupal, Wordpress) that enables a design of websites without learning web technologies, we reduce the gap of learning semantic web technologies by developing a framework that assists developers designing cross-domain SWoT applications. We design and implement SWoTSuite\footnote{\url{http://sensormeasurement.appspot.com/}}, a suite of tools for prototyping SWoT applications. We first present an architecture of SWoTSuite and then an application development process using it.

\section{SWoTSuite architecture}\label{sec:architecture} 
In the following, we describe an architecture of SWoTSuite and its components.

\fakeparagraph{Semantic annotator} It interacts with IoT devices that return data in heterogeneous formats. It converts sensors metadata in an unified description to provide further reasoning to overcome heterogeneity issues, similar to approach adopted by Semantic Sensor Observation Service~(SemSOS)~\cite{semsos}.  The sensor metadata is semantically annotated according to the M3 taxonomy~\cite[p.~93]{gyrard:tel-01217561}(an extension of W3C Semantic Sensor Network~(SSN) ontology) that covers the limitations of SSN~\cite{M3ontology}.

\begin{figure}[!h]
\centering
\includegraphics[width=0.77\linewidth]{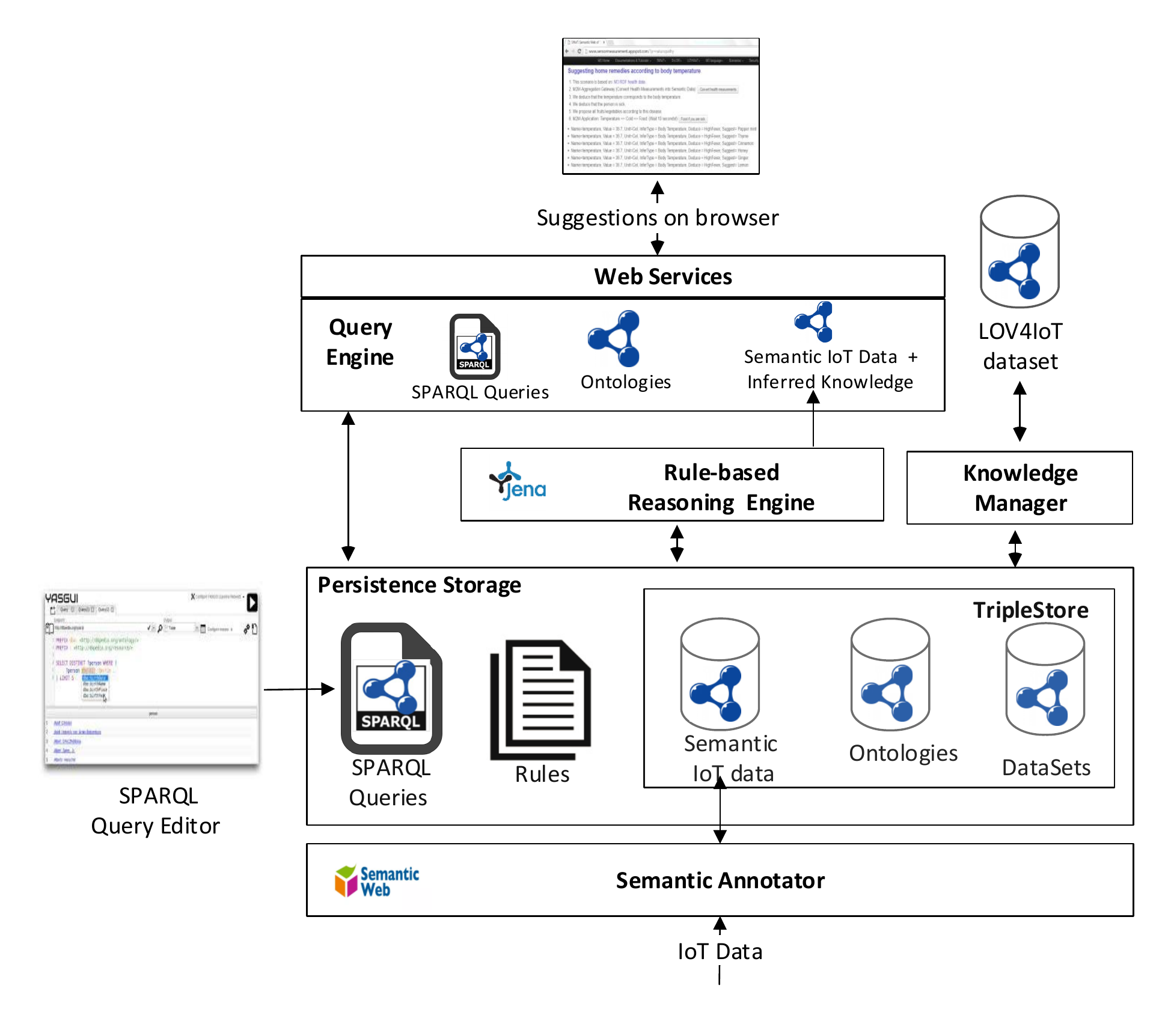}
\caption{SWoTSuite architecture}
\label{swotarchitecture} 
\end{figure}

\fakeparagraph{Persistent storage} It stores the M3 ontologies, M3 datasets\cite{gyrard:tel-01217561}, M3 rules\cite{gyrard:tel-01217561}, and unified sensor data received from the annotator. Moreover, it stores pre-written M3 compatible SPARQL queries to assist developers.  The current implementation of persistent storage uses a triple store to store M3 ontologies, M3 datasets, unified sensor data received from the converter. The M3 rules and SPARQL queries are stored as files. 

\fakeparagraph{Knowledge manager}  It is responsible for indexing, designing, reusing, and combining  domain-specific knowledge that further update knowledge in the persistent storage. We have been building datasets to reference, classify, and reuse domain-specific knowledge that we call Linked Open Vocabularies for the 
Internet of Things~(LOV4IoT)\footnote{\url{http://sensormeasurement.appspot.com/?p=ontologies}}. The LOVIoT provides domain ontologies, datasets, and rules that could be reused to design cross-domain SWoT applications. 

\fakeparagraph{Reasoning engine} It infers high-level knowledge using a Jena inference engine and M3 rules. The M3 rules are a set of rules that are extracted from LOV4IoT and redesigned to be compliant with the M3 taxonomy. 

\fakeparagraph{Query engine} It executes queries and provides smart suggestions. The query engine has been implemented using ARQ, a SPARQL processor for Jena. The query engine loads the M3 ontologies, M3 datasets, knowledge deduced from the reasoning engine, and executes SPARQL queries in order to provide suggestions.

\subsection{Demonstration}\label{sec:demo} 
For demonstration purposes, we take an application that combines data produced by a body thermometer with other domains such as healthcare and food knowledge bases. It suggests some preventive measures, such as home remedies when a fever is detected~(e.g., drink orange or lemon juice because they contain Vitamin C) or consulting a doctor immediately in case of severity, that can be acted upon by patients. 

An application development using SWoTSuite consists of the following steps. The detailed steps can be found at URL \url{http://sensormeasurement.appspot.com/?p=end_to_end_scenario}. The recorded video demonstrating application development process using SWoTSuite is available at URL \url{http://tinyurl.com/zs9flrt}.

\fakeparagraph{Step~1:~Generating an IoT application template}
It selects a pre-defined template from the persistent storage. The template consists of ontologies, datasets, rules, and SPARQL queries that are required  to build a SWoT application.  To select a template, developers provide a set of sensors (e.g., body thermometer) and the related application domain (e.g., healthcare) information. SWoTSuite queries the persistent storage with this information and returns templates to select. 

\begin{figure}[!h]
\centering
\includegraphics[width=1.1\linewidth]{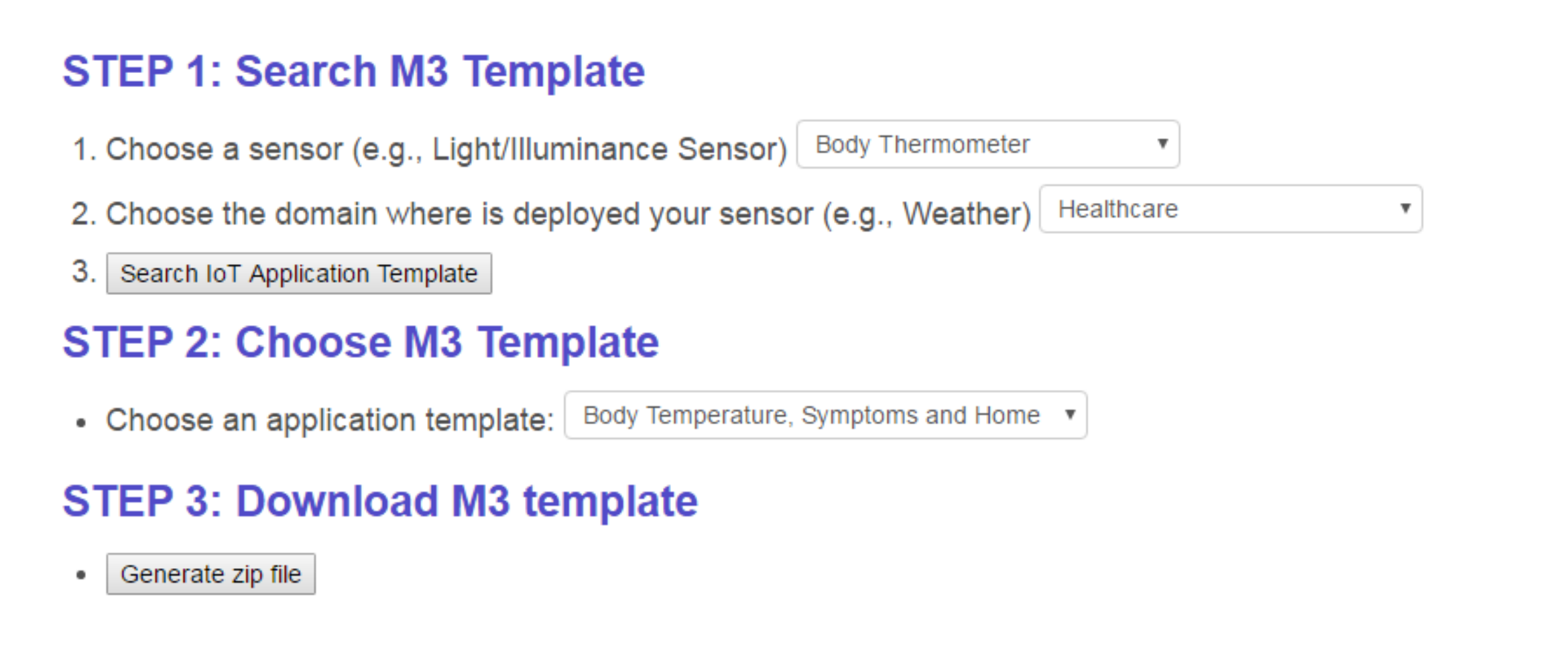}
\caption{The user interface  to generate IoT application template. The user interface is available to use at URL: \protect\url{http://sensormeasurement.appspot.com/?p=m3api}}
\label{fig:templategeneration}
\end{figure}

Figure~\ref{fig:templategeneration} shows this multi-step process. In the first step, users provide sensors and and application domain information. With these information, SWoTSuite searches application template and allows developers to select an appropriate template. In the next step, an appropriate template is available to further use. 

\begin{figure}[!h]
\centering
\includegraphics[width=1.0\linewidth]{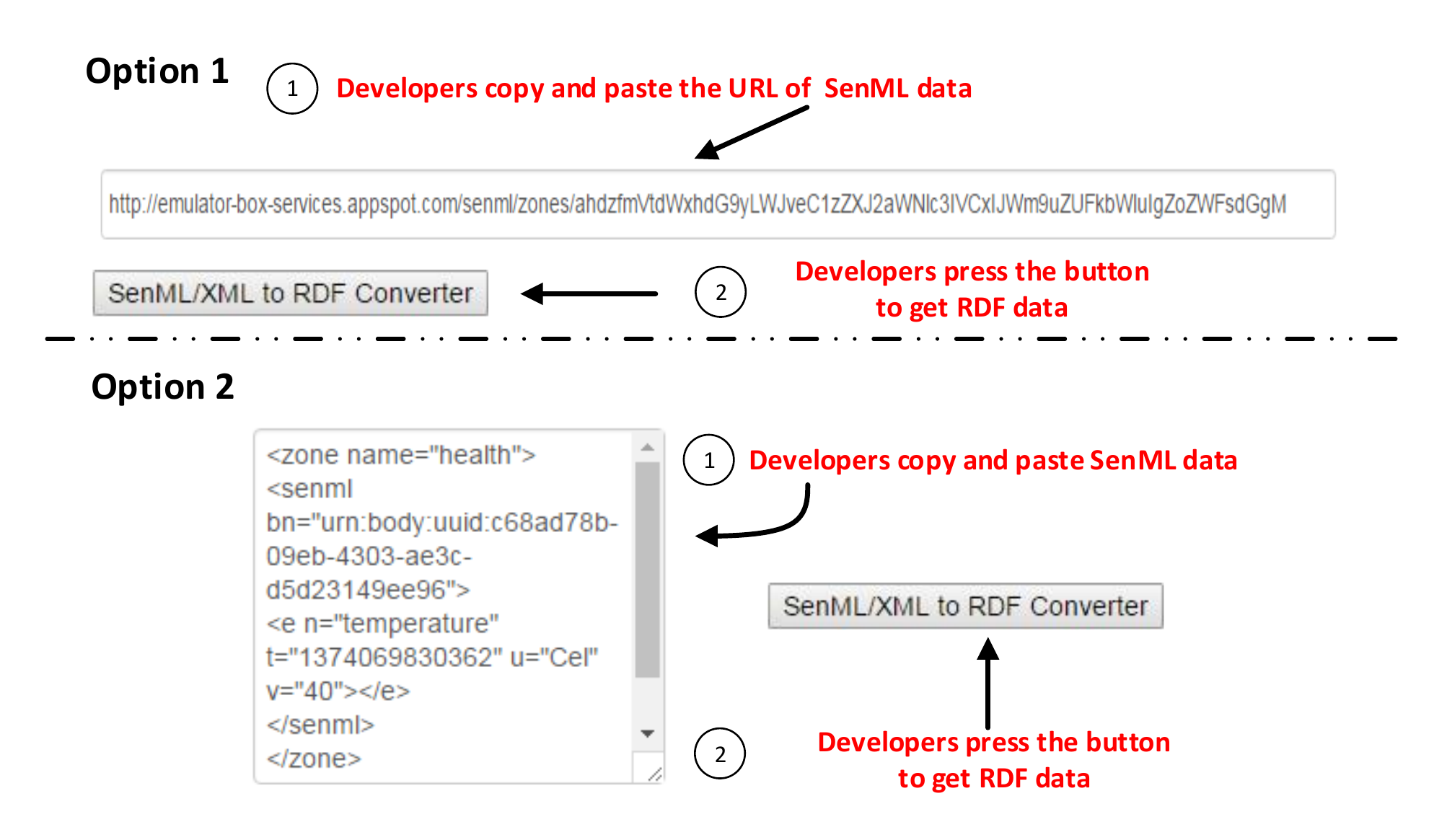}
\caption{Semantic Annotator user interface to generate RDF/XML data. The user interface is available to use at URL: \protect\url{http://www.sensormeasurement.appspot.com/?p=senml_converter}}
\label{fig:annotator}
\end{figure}

\fakeparagraph{Step~2:~Semantically annotating data} Developers provide SenML data to the semantic annotator to get the RDF/XML data. We provide two options for interacting with the semantic annotator, as shown in Figure~\ref{fig:annotator}. Option~1 allows the developers to enter a URL of data source; in  option~2, the developers enter SenML data manually and press the converter button to get annotated data. Currently, we have been working on different ways of annotating sensor data. One of ways is tools like Kino~\cite{kino} does. The kino accepts documents via a special interface\footnote{\url{https://www.youtube.com/watch?v=12R81HrlAF8}}.   

\fakeparagraph{Step~3:~Executing the reasoning engine} It deduces new knowledge from annotated IoT data. To perform this task, developers are guided through Linked Open Rules available in the persistent storage. Listing~\ref{rules} shows a small code snippet of a health-care domain.
An example of a rule is ``if body temperature is between  37.5$^{\circ}$C--39$^{\circ}$C, then high fever''. 

Listing~\ref{rules} shows a small code snippet of a health-care domain. This rule deduces a condition based on the value of the body thermometer data. Moreover, the developers are guided through Java code to load rules. Listing~\ref{javarulesload} illustrates a small code snippet to load the rules into the Jena reasoning engine.

\lstset{ 
     emph={greaterThan, lessThan}, emphstyle={\color{blue}\bfseries}, 		
		caption={Code snippet of rules. This rule indicate -- if body temperature is greater than 37.5$^{\circ}$C and lower than 39$^{\circ}$C then highfever. The language keywords are printed in {\color{blue} {\texttt{\textbf{blue}}}}. For a full Listing of rules, refer the URL: \protect\url{http://sensormeasurement.appspot.com/tutorial/M3ApplicationTutorial.zip} }, escapechar=\#,  label=rules}	
\begin{lstlisting}
[
HighFever: 
    (?measurement rdf:type m3:BodyTemperature)
    (?measurement m3:hasValue ?v)
		greaterThan(?v,37.5) 
		lessThan(?v,39.0) 
		->
    (?measurement rdf:type health-dataset:HighFever)
]
\end{lstlisting} 

\lstset{language=Java, caption={Code snippet to load the rules files into the Jena reasoning engine. For a full Listing of rules, refer the URL: \protect\url{http://sensormeasurement.appspot.com/tutorial/M3ApplicationTutorial.zip}}, escapechar=\#, label=javarulesload} 
\begin{lstlisting}
public Model executeReasoningEngine(String JenaRuleFile)
{
	//Load the Jena rules to deduce meaningful knowledge.
	Reasoner reasoner = new GenericRuleReasoner(Rule.rulesFromURL(JenaRuleFile));// read rules
	
	reasoner.setDerivationLogging(true);
	//Apply the Jena Reasoning Engine
	
	InfModel inf = createInfModel(reasoner, model);
	
	//Return Data with more meaningful knowledge.
	return inf;
}
\end{lstlisting}

\fakeparagraph{Step~4:~Executing the query engine} It executes SPARQL queries and provides suggestions. We guide  developers to use ARQ, a SPARQL query processor for Jena. The query engine loads the M3 ontologies, datasets, annotated data~(Step~1), derived suggestions~(Step~3), and SPARQL queries. Listing~\ref{query} illustrates the code snippet of a SPARQL query for the naturopathy scenario. The SPARQL query 
derives recommendations over M3 ontologies, M3 ontologies, domain-specific Health and Medicine datasets.

\lstset{ 
     emph={SELECT, DISTINCT, WHERE, OPTIONAL, FILTER, LANGMATCHES, LANG, str}, emphstyle={\color{blue}\bfseries}, 		
		caption={A code snippet of a SPARQL query for deriving suggestions. The language keywords are printed in {\color{blue} {\texttt{\textbf{blue}}}}. For a full Listing of rules, refer the URL: \protect\url{http://sensormeasurement.appspot.com/tutorial/M3ApplicationTutorial.zip} }, escapechar=\#,  label=query}	
\begin{lstlisting}
SELECT DISTINCT  ?measurement ?name ?value ?unit ?inferType ?suggest  WHERE{
	?measurement m3:hasName ?name.
	?measurement m3:hasValue ?value.
	?measurement m3:hasDateTimeValue ?time.
	?measurement m3:hasUnit ?unit.
	
	?measurement rdf:type ?inferTypeUri.
		OPTIONAL { 
			?inferTypeUri rdfs:label ?inferType. 
			FILTER(LANGMATCHES(LANG(?inferType), "en"))
		}
		
		OPTIONAL {
			?measurement rdf:type ?deduceUri.
			?deduceUri rdfs:label ?deduce.
			FILTER(LANGMATCHES(LANG(?deduce), "en"))
			FILTER(str(?deduceUri) != str(m3:Measurement) )
			FILTER(str(?deduceUri) != str(?inferTypeUri) )
		} 
		
		OPTIONAL{
			?resUri  m3:hasRecommendation ?deduceUri. 
			?resUri rdf:type ?typeRecommendedUri.  
			?resUri rdfs:label ?suggest.
			FILTER(LANGMATCHES(LANG(?suggest), "en"))
		}	
}
\end{lstlisting}

\fakeparagraph{Step~5:~Displaying results} It receives results from the query engine~(Step~4) and displays derived knowledge and suggestions on a Web browser. The user interface is implemented with HTML5, CSS3, JavaScript and AJAX technologies to query SWoTSuite web services. 
Other functionality can be achieved in this layer such as controlling actuators, sending alerts to a mobile device.

\begin{figure}[!h]
\centering
\includegraphics[width=1.05\linewidth]{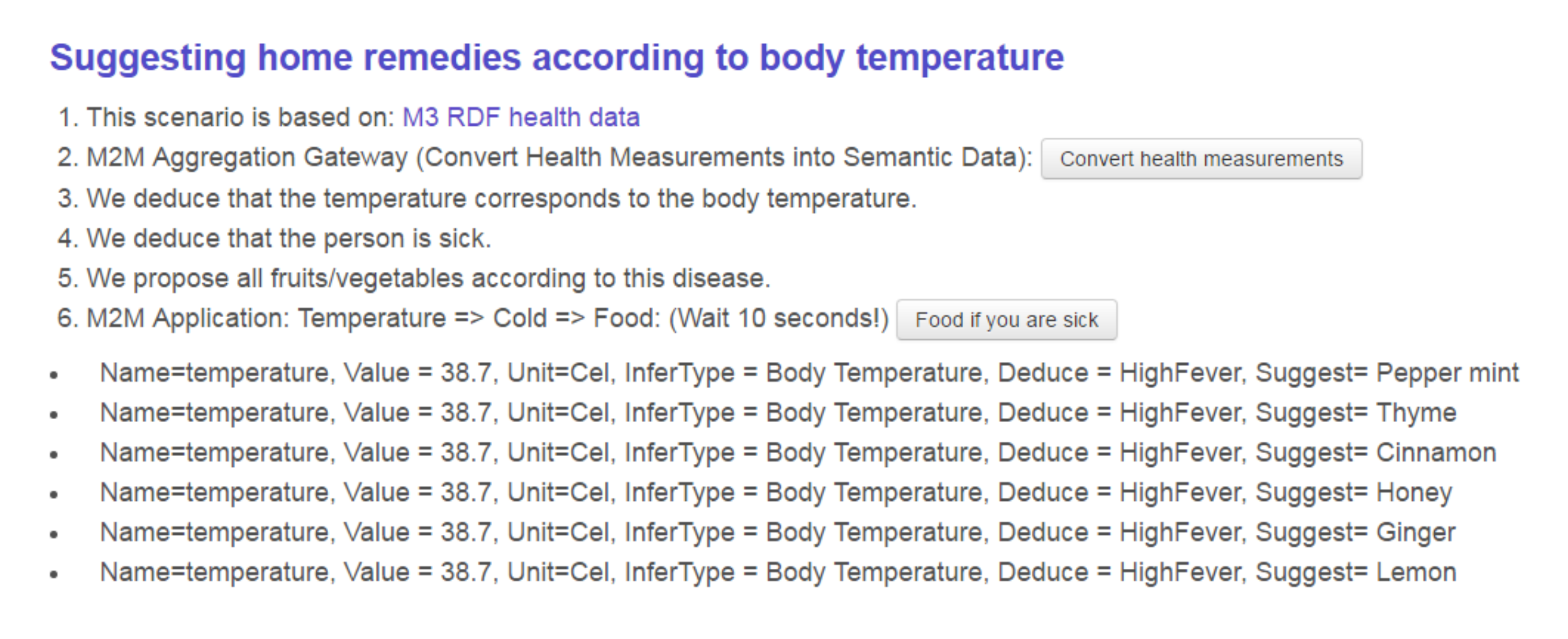}
\caption{The naturopathy scenario suggesting home remedies when a fever is deduced. The user interface is available to use at URL: \protect\url{http://www.sensormeasurement.appspot.com/?p=naturopathy}}
\label{fig:userinterface}
\end{figure}

For our naturopathy use case, SWoTSuite deduces a fever suggests home remedies as shown in Figure~\ref{fig:userinterface}. 
The current version is limited to a browser for suggestions. We have been working an Android application that receives suggestions from SWoTSuite and display is in more presentable manner. 

\fakeparagraph{Evaluation of SWoTSuite} We have evaluated our framework on three aspects: the performance of the semantic annotator given a size of data~\cite[p.~85]{gyrard:tel-01217561}, the performance of the reasoner given a number of rules~\cite[p.~86]{gyrard:tel-01217561}, and the performance of different tasks according to a size of data~\cite[p.~87]{gyrard:tel-01217561}. 

\subsection{Summary and Future work}  
In this paper, we demonstrate SWoTSuite, a toolkit for prototyping SWoT applications. It hides the use of semantic web
technologies as much as possible to avoid the burden of designing SWoT applications that involves designing ontologies, annotating sensor data,
and using reasoning mechanisms to enrich data. We take a hello world naturopathy application as an example and demonstrate an application
development process using SWoTSuite. 

We believe that the naturopathy application is too small to demonstrate the capabilities and usefulness of 
SWoTSuite. Therefore, our immediate plan is to derive more concrete scenarios. We see that the possible source of scenarios is CityPulse project usecases\footnote{\url{http://www.ict-citypulse.eu/page/content/smart-city-use-cases-and-requirements}}. Second future work will be on
evaluating SWoTSuite framework on two aspects: (1) We are planning to evaluate this framework on \emph{development efforts}~\cite{patel:hal-00788366, DBLP:journals/corr/PatelLB16}. (2) We are going to evaluate the \emph{usability} of SWoTSuite with real users through a Google form\footnote{The Google form is available at\url{https://docs.google.com/forms/d/e/1FAIpQLSd1XQoTUdUg48-Ckc25iw4BIpTUPu3SxCOWLUcGdLOxD0kM0Q/viewform}}  at our ISWC 2016 tutorial~\cite{iswctutorial}. The web link of this tutorial is available at URL\footnote{\url{http://sensormeasurement.appspot.com/?p=ISWC2016Tutorial}}. 
\bibliographystyle{splncs}

\end{document}